\documentclass[a4paper,
               keeplastbox,   
               ]{jacow}
\usepackage{pdfpages,multirow,ragged2e} %
\usepackage[Symbol]{upgreek}
\usepackage{color}  
\usepackage{hyperref}
\usepackage{graphics}
\usepackage{subcaption}
\hypersetup{
    colorlinks=false,       
    linkbordercolor=green,          
    citecolor=black,        
    filecolor=magenta,      
    urlcolor=blue           
}

\makeatletter%
	\ifboolexpr{bool{xetex}}
	 {\renewcommand{\Gin@extensions}{.pdf,%
	                    .png,.jpg,.bmp,.pict,.tif,.psd,.mac,.sga,.tga,.gif,%
	                    .eps,.ps,%
	                    }}{}
\makeatother

\ifboolexpr{bool{xetex} or bool{luatex}} 
 {}                                      
 {\usepackage[utf8]{inputenc}}           

\usepackage[USenglish]{babel}

\ifboolexpr{bool{jacowbiblatex}}%
 {%
  \addbibresource{jacow-test.bib}
  \addbibresource{biblatex-examples.bib}
 }{}
\begin{document}

\newcommand{\pt}{\ensuremath{p_\mathrm{T}}}
\newcommand{\pT}{\pt}
\newcommand{\pp}{\ensuremath{\mbox{p}\mbox{p}~}}
\newcommand{\Minv}{\ensuremath{M_\mathrm{inv}}}
\newcommand{\Qinv}{\ensuremath{Q_\mathrm{inv}}}

\title{Event shapes and jets in $e^{+}e^{-}$ and pp collisions}

\author{M. Sas\textsuperscript{1,2}, J. Schoppink\textsuperscript{1}\\
		\textsuperscript{1}Institute for Subatomic Physics, Utrecht University/Nikhef, Utrecht, Netherlands\\
		\textsuperscript{2}Physics Department, Yale University, New Haven CT, U.S.A.\\
		\today}
	
\maketitle

\begin{abstract}
In high energy particle collisions the shape of the event, i.e. the relative distribution of particles in momentum space, is often used to try to select events with certain topologies. It is claimed that an event shape observable like transverse sphericity is able to discriminate between jet-like events and events that are dominated by soft production from the underlying event.

In this paper we investigate the relationship between the shape of the event and the number of jets found in the respective event for both  $e^{+}e^{-}$ and pp collisions using the PYTHIA model. In $e^{+}e^{-}$ collisions, we find that the transverse sphericity of the event can be used effectively to either enhance or suppress the fraction of jets found in the selected sample, and can even discriminate between single, two, and multi-jet topologies. However, contrary to current literature, we find that in pp collisions this does not hold. It is shown that the transverse sphericity as well as the particle multiplicity is sensitive to the number of multi-parton interactions.
\end{abstract}

\section{Introduction}

Transverse sphericity is an event shape observable that has been used throughout the particle physics community as a way to characterize the configuration of the momentum vectors of the final state particles as either pencil-, sphere-like, or anything in between. Pencil-like events are typically associated with events that have a di-jet structure, and sphere-like events with those containing mostly soft processes and an absence of jets \cite{ALICE_1,ALICE_2}.

Historically, quantifying the shape of the event involved calculating the thrust axis, i.e. the axis which maximizes the inner product of the final-state particle momentum vectors. As such, the thrust axis aligns with the average momentum of the particles and is a good approximation for the main production axis of the event, especially for $e^{+}e^{-}$ collisions \cite{Ellis:1980wv,Abe:1994mf,Abdallah:2003xz,Achard:2004sv,Weinzierl:2009ms}, where to leading order there are two partons produced with opposite momenta. For this collision system pencil-like events should correspond to a di-jet structure, and spherical events to multi-jet topologies \cite{Dasgupta:2003iq,Abbiendi:2004qz,Kluth:2006bw,Gehrmann-DeRidder:2007nzq}.

This paper addresses whether this holds for \pp collisions at RHIC and top LHC energies, where there are more sub-leading processes in addition to the initial hard scattering \cite{201148,Banfi:2010xy}. This paper presents the results of an investigation on how the transverse sphericity $(S_{\mathrm{T}})$ correlates with jet production, using simulated $e^{+}e^{-}$ and \pp collisions. In addition, correlations between $(S_{\mathrm{T}})$ and other event characteristics, such as leading parton \pT, particle multiplicity, and number of multi-parton interactions are studied.

\section{Analysis method}

The analysis presented in this paper is performed using simulated $e^{+}e^{-}$ and \pp collisions, using the PYTHIA8.1 event generator \cite{pythia}, with the settings listed in Table \ref{tab:DatasetsEventShapes}. The center-of-mass energy of the $e^{+}e^{-}$ dataset corresponds to the mass of the $Z$ boson, which is also the energy at which the LEP collider operated. For the pp datasets the beam energies are chosen to be the same as the nominal operating energy at the RHIC and the LHC in Run 2, both using the Monash 2013 tune. Furthermore, the pp datasets are generated twice, once with multi-parton interactions (MPI) turned on, and once with MPI turned off. This will be used to study the sphericity distributions with and without MPI in pp collisions. Tracks with $\pT>100~$MeV$/c$ and $|\eta|<1$ are selected for the analysis, and in the case of the $e^{+}e^{-}$ dataset only the hadronic final states are considered.

The transverse sphericity is calculated using
\begin{equation}
S_{\mathrm{T}} = \frac{2 \lambda_{2}}{\lambda_{2}+\lambda_{1}},
\end{equation}
where $\lambda_{1}$ and $\lambda_{2}$, with $\lambda_{1} > \lambda_{2}$, are the two eigenvalues of the transverse momentum matrix $S_{xy}^{L}$, which is given by
\begin{equation}
S_{xy}^{L} = \frac{1}{\sum_{i} p_{T,i}} \sum_{i} \frac{1}{p_{T,i}}
\begin{bmatrix}
p_{x,i}^{2} & p_{x,i}p_{y,i} \\
p_{y,i}p_{x,i} & p_{y,i}^{2}
\end{bmatrix},
\end{equation}
where $p_{T,i}$, $p_{x,i}$, and $p_{y,i}$, are the components of the momentum vectors of particle $i$ in transverse, $x$, and $y$ direction, respectively.\\
These equations essentially project all the particles of the event onto the $x-y$ plane and determine the eigenbasis of the transverse momentum matrix. As such, $S_{\mathrm{T}}$ is sensitive to the relative orientation of the momentum vectors, where $\lambda_{2}=0$ is for pencil-like events, i.e. $S_{\mathrm{T}}\sim0$, and $\lambda_{1} \sim \lambda_{2}$ for sphere-like events, i.e. $S_{\mathrm{T}}\sim1$. 

Furthermore, the number of jets contained in each event is obtained by employing the FastJet package \cite{fastjet}. We chose to use the anti$-k_{t}$ algorithm with a jet radius of $R=0.4$ and minimum jet energy of $E=10~$GeV. As this study is purely based on model calculations, it is chosen to only include statistical uncertainties that are proportional to the number of events as generated in the respective dataset. In all cases the datasets are large enough to lead to statistically significant conclusions.

\begin{table}[htb]
\centering
	\begin{tabular}{l|l|l} \toprule
	\textbf{System} & \textbf{$\sqrt{s}$} & \textbf{PYTHIA8.1 settings}\\
	\hline
	$e^{+}e^{-}$ & $91~$GeV & $Z$ decay to quarks\\
	\hline
	\pp & $200~$GeV & Monash tune, MPI-ON\\
	\hline
	\pp & $200~$GeV & Monash tune, MPI-OFF\\
	\hline
	\pp & $13~$TeV & Monash tune, MPI-ON\\
	\hline
	\pp & $13~$TeV & Monash tune, MPI-OFF\\
	\midrule
	\bottomrule
	\end{tabular}
\caption{Data sets used.}
\label{tab:DatasetsEventShapes}
\end{table}

\section{Results}

\begin{figure*}[htb]
\centering
  \includegraphics[width=0.44\linewidth]{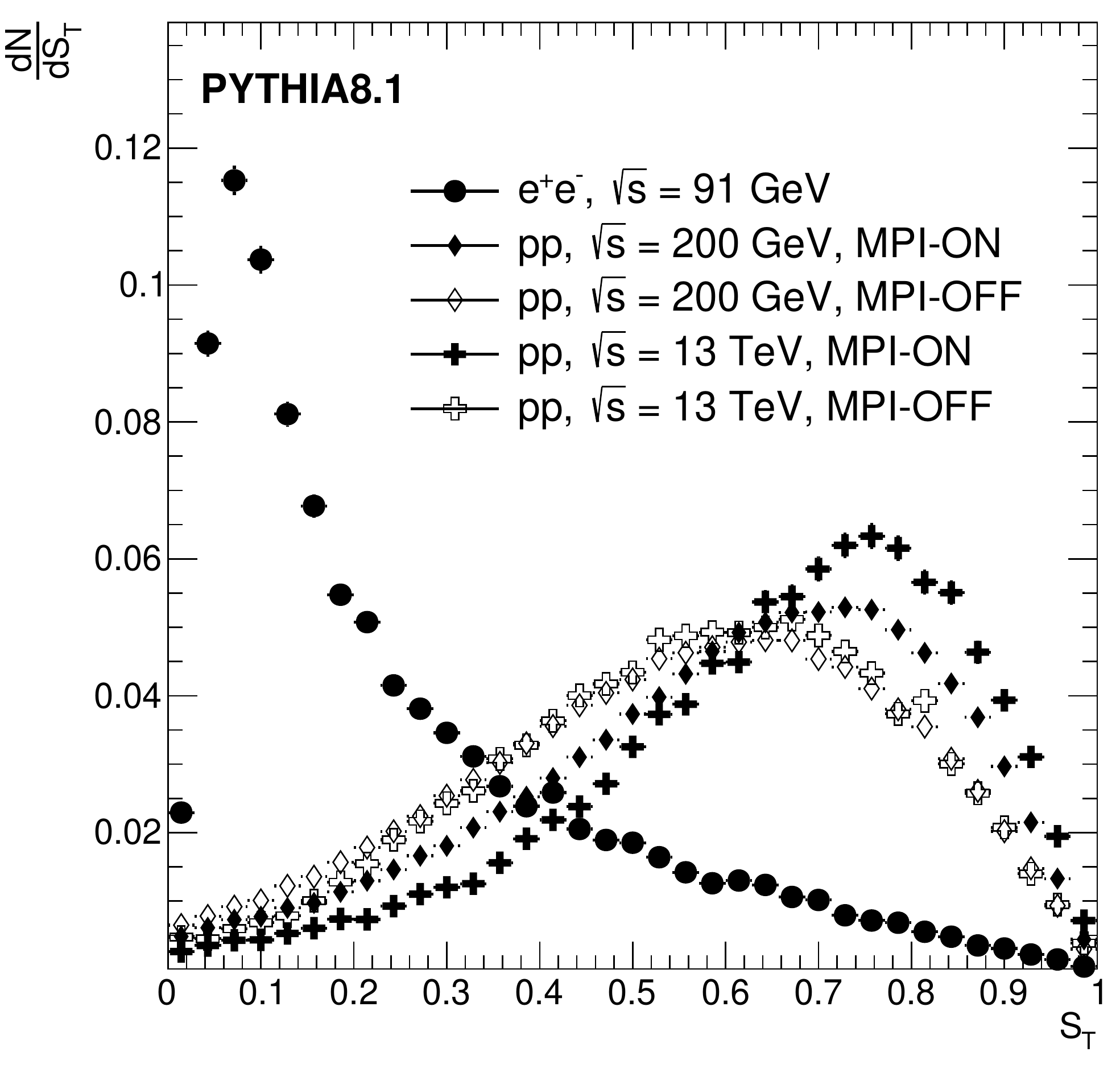}
  \includegraphics[width=0.44\linewidth]{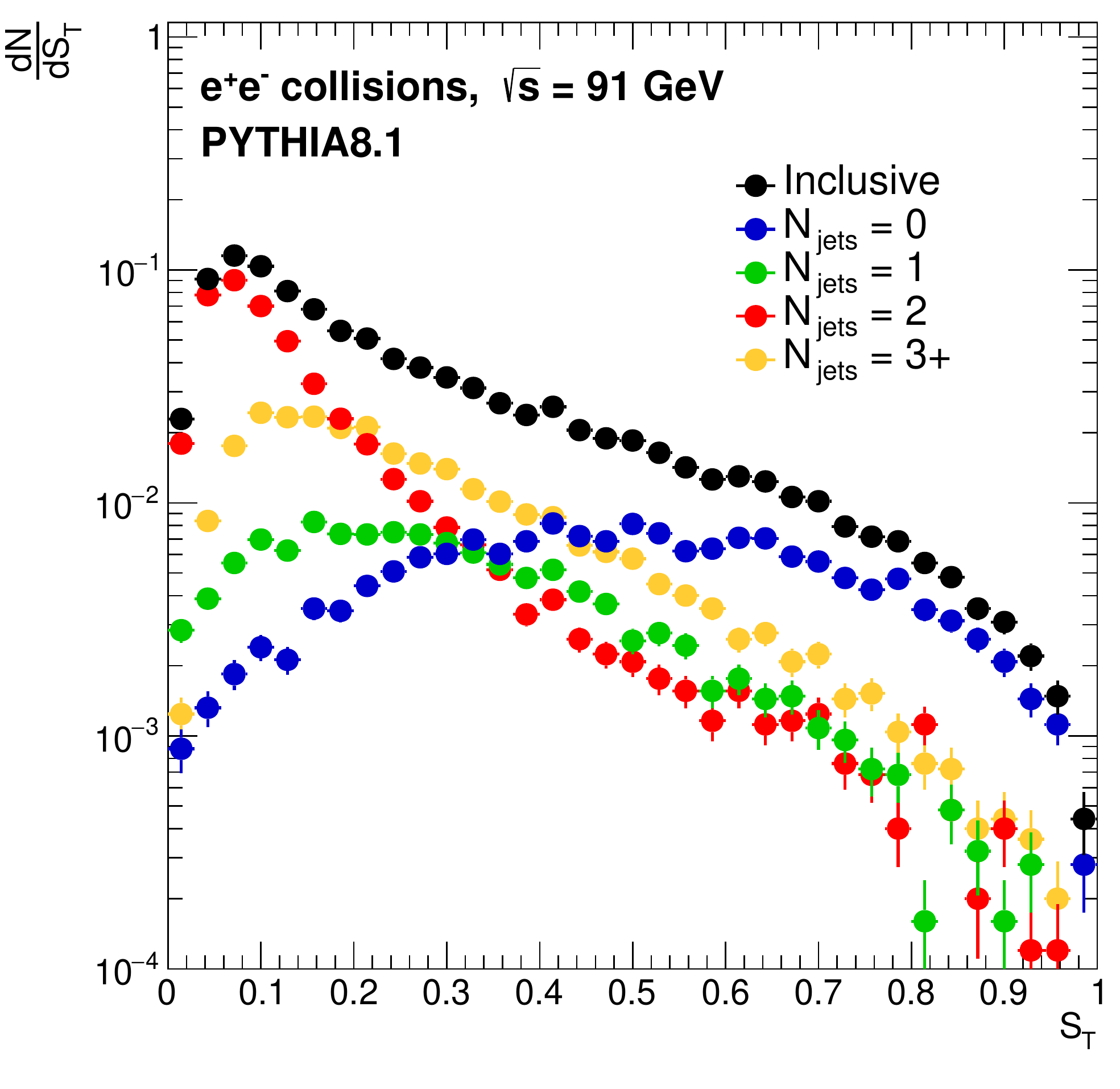}\\
  \includegraphics[width=0.44\linewidth]{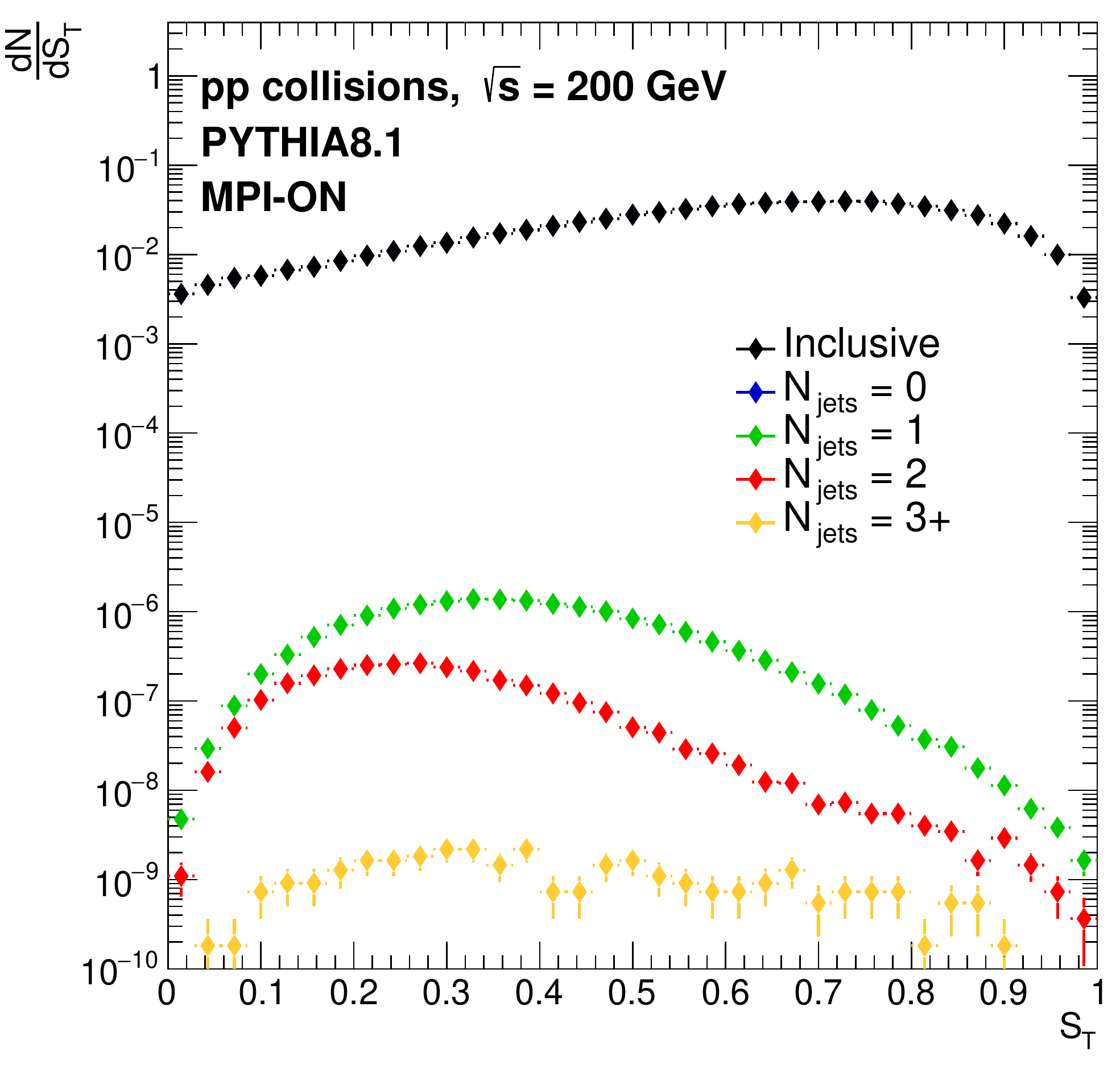}
  \includegraphics[width=0.44\linewidth]{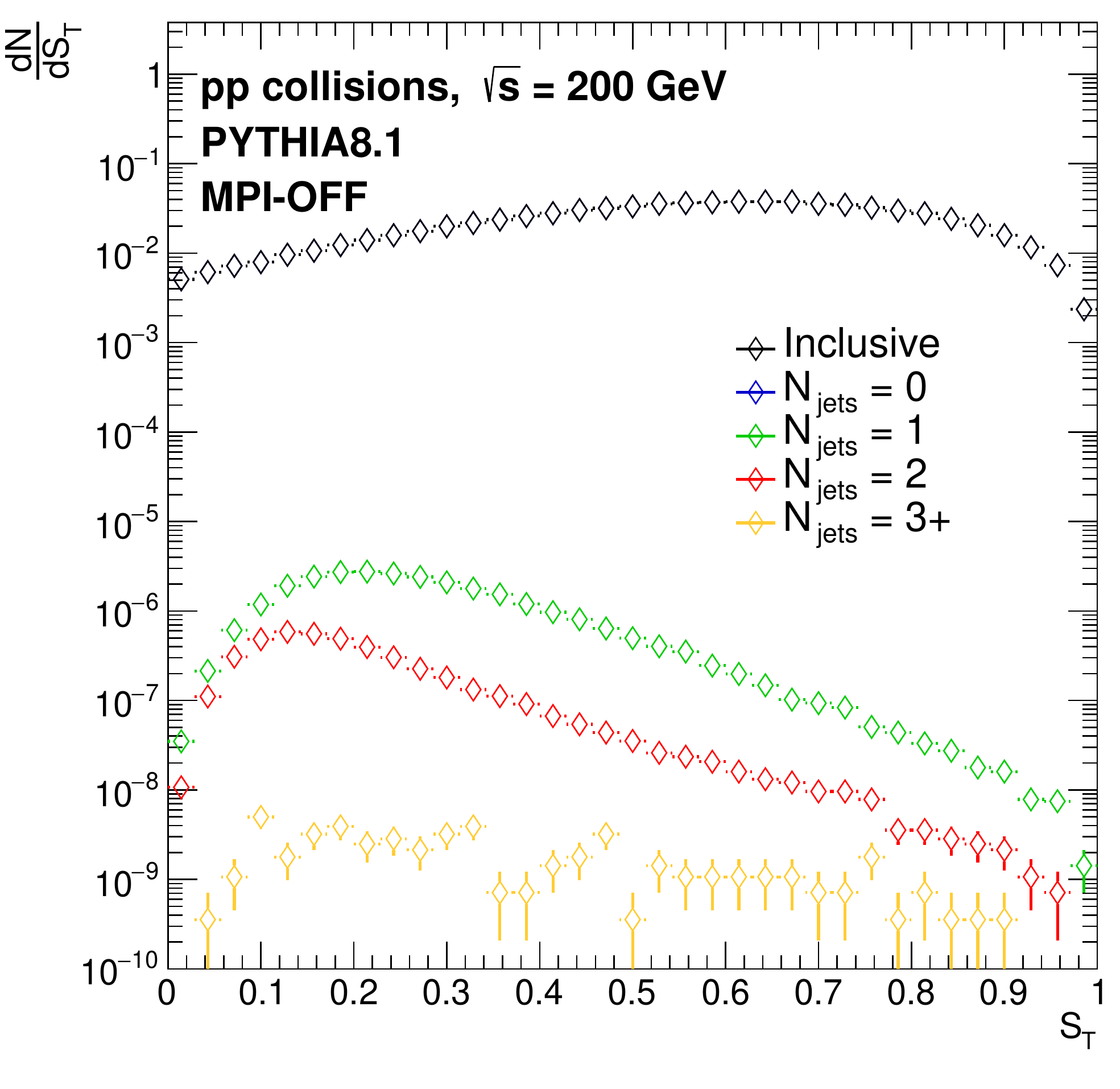}\\
  \includegraphics[width=0.44\linewidth]{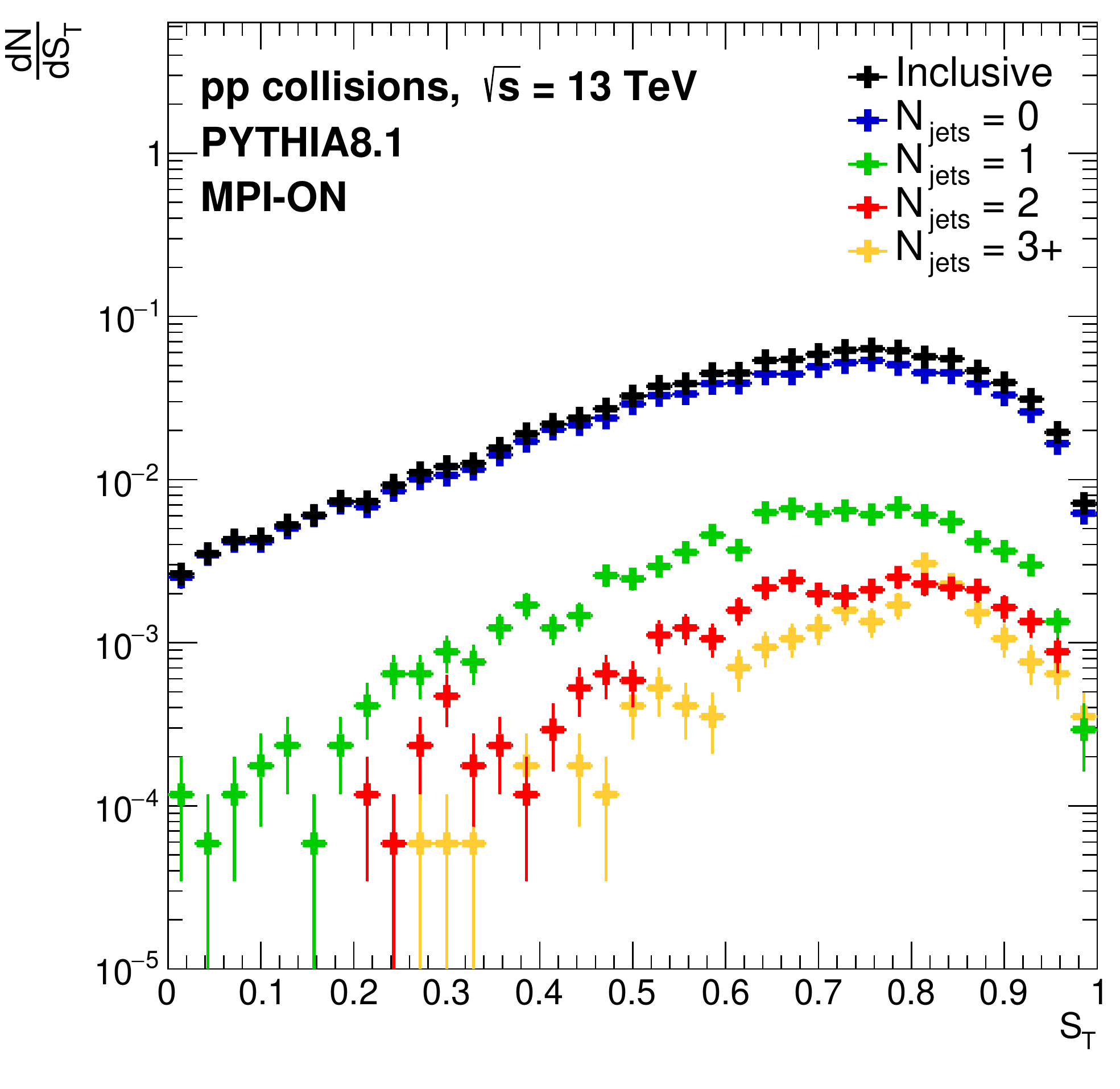}
  \includegraphics[width=0.44\linewidth]{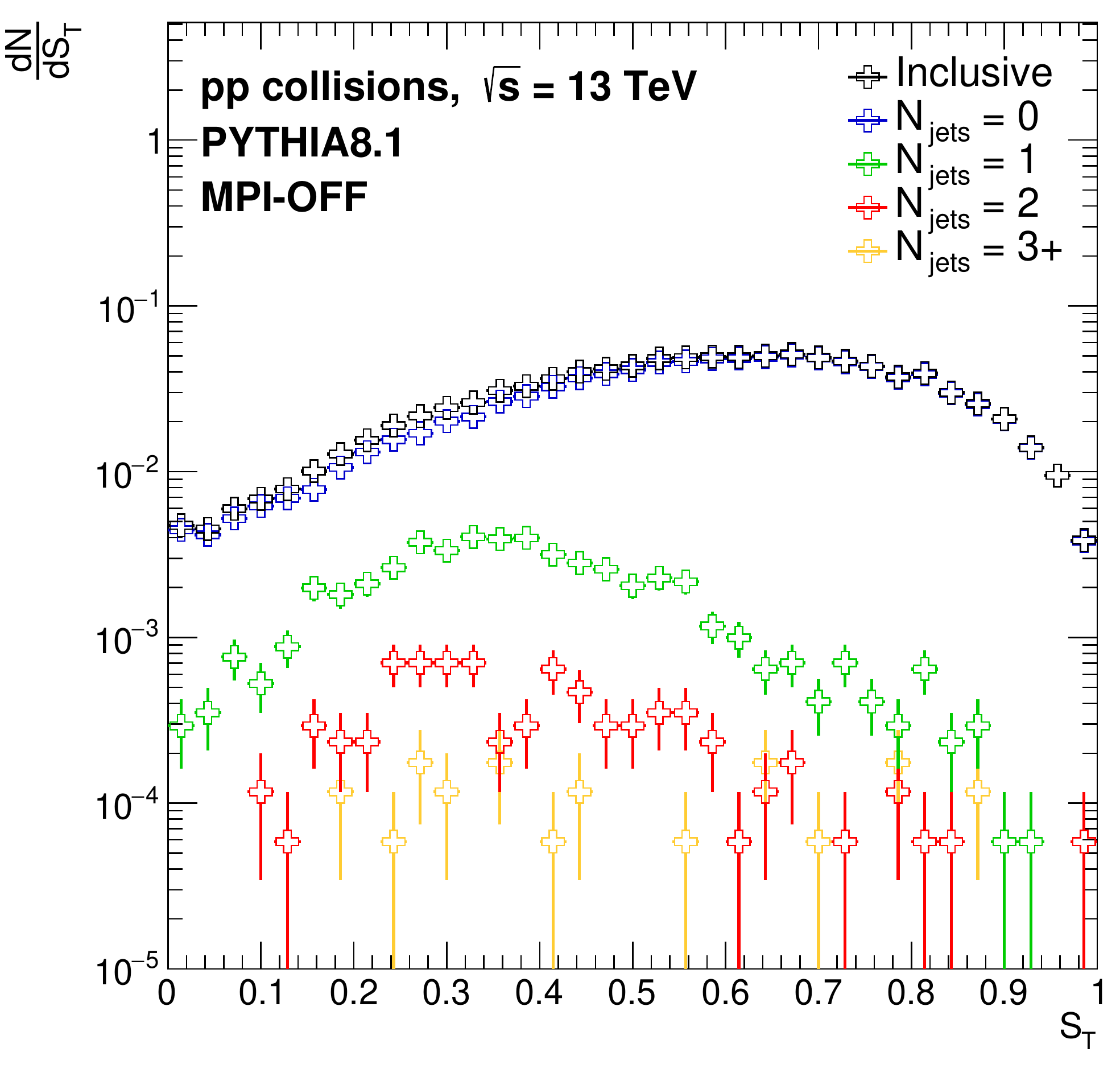}\\
  \caption{Transverse sphericity distributions $(S_{\mathrm{T}})$ for simulated $e^{+}e^{-}$ and \pp collisions (top left), and differentiated for the amount of jets reconstructed in the events $(N_{\mathrm{jets}}=0,1,2,3+)$ for each respective dataset. The inclusive distributions are normalized to unity. The inclusive and $N_{\mathrm{jets}}=0$ distributions for the pp collisions dataset at $\sqrt{s}=200~$GeV are indistinguishably close.}
  \label{fig:EventShape_all}
\end{figure*}

The distributions of the transverse sphericity $S_{\mathrm{T}}$ for the respective datasets as well as for events with different numbers of reconstructed mid-rapidity jets are shown in Fig. \ref{fig:EventShape_all}. The top left figure shows first of all that the $S_{\mathrm{T}}$ distribution for $e^{+}e^{-}$ collisions has a maximum around $S_{\mathrm{T}}\sim0.1$, with a long tail towards higher values. This is consistent with the picture that lepton collisions, to leading order, produce two outgoing quarks that fragment into final state particles. The $S_{\mathrm{T}}$ distributions for the \pp datasets have a mean around $S_{\mathrm{T}}\sim0.75$ for MPI-ON, and $S_{\mathrm{T}}\sim0.6$ for MPI-OFF. Clearly, \pp collisions produce a much broader distribution with a larger mean. There is no abundance of events with a pencil-like configuration, which is consistent with the idea that \pp collisions involve many more processes compared to the ``cleaner'' $e^{+}e^{-}$ collisions, tending to be more spherical due to the overall higher particle multiplicities produced by a larger number of independent processes.

Figure \ref{fig:EventShape_all} shows the $S_{\mathrm{T}}$ distributions for the number of mid-rapidity jets reconstructed in the events, for $N_{\mathrm{jets}}=0,1,2,3+$. In the top right figure, the simulated $e^{+}e^{-}$ collisions are dominated by 2-jet events at lower $S_{\mathrm{T}}$. For slightly higher values, $N_{\mathrm{jets}}=3+$ contributes the most, after which for $S_{\mathrm{T}}>0.5$ the events with $N_{\mathrm{jets}}=0$ take over and dominate the contribution to the inclusive sample. Single jet events exhibit a broad distribution that have a maximum at lower values of $S_{\mathrm{T}}$. For \pp collisions, strikingly, the entire sample is dominated by events that do not have a jet that passes the selection criteria. This is consistent with the notion that, even though the collision energy is much higher, \pp collisions have on average significantly fewer high momentum jets in the final state compared to $e^{+}e^{-}$ collisions, as the rapidity distribution is wider and tracks with $|\eta|<1$ are selected. For the pp dataset with MPI-ON at $\sqrt{s}=13~$TeV, contrary to the same distributions found at $\sqrt{s}=200~$GeV, the single and multi-jet events are more pronounced for higher values of $S_{\mathrm{T}}$, while for the datasets with MPI-OFF the events containing a jet have a lower sphericity value. This indicates that multi-parton interactions do not play a major role at RHIC energies.

To better quantify the effect of making a selection in $S_{\mathrm{T}}$ on the number of jets in the sample, the respective distributions for the highest $10\%$ of the most pencil- and sphere-like events are integrated and the fraction of each to the inclusive distribution is calculated. The results of this calculation are given in Table \ref{tab:EventShape_jets_fractions_10}. First, it shows that $e^{+}e^{-}$ collisions have jets in all but $16.1\%$ of the events, while this fraction of non-jet events rises to $85.5\%$ and $93.3\%$ for \pp collisions at $\sqrt{s}=13~$TeV with MPI-ON and MPI-OFF, respectively. Most of the $e^{+}e^{-}$ collisions contain $2$ jets, followed by $3+$ jets. For \pp collisions the majority of events do not contain a jet, followed by single-jet events. These events are most likely events where one of the two jets is outside of the detector acceptance, i.e. the majority of the higher $\pT$ constituent tracks have $|\eta|>1$. In $e^{+}e^{-}$ collisions the fraction of events containing $N_{\mathrm{jets}}=2$ increases from $44.9\%$ for all events to $83.3\%$ for the $10\%$ most pencil-like events. It decreases from $3.2\%$ to $1.3\%$ for pp collisions at $\sqrt{s}=13~$TeV with MPI-ON and slightly increases for the case with MPI-OFF. For the $10\%$ most sphere-like events, the fraction of events without a jet in $e^{+}e^{-}$ collisions increases from $16.1\%$ for all events to $58.8\%$ in the most spherical-like events, followed by a sizeable contribution from $N_{\mathrm{jets}}=3+$. Selecting on spherical events in \pp collisions doesn't change the fractions of events that have a jet like it does for $e^{+}e^{-}$ collisions, but shows an increase in $N_{\mathrm{jets}}=2$ from $3.2\%$ to $4.4\%$ for pp collisions at $\sqrt{s}=13~$TeV with MPI-ON. As jets are so rare in pp collisions at $\sqrt{s}=200~$GeV, none of the selections have a significant effect on the $N_{\mathrm{jets}}$ fractions. These results indicate that for $e^{+}e^{-}$ collisions a selection on the sphericity is able to enhance a specific amount of jets within that sample, while for \pp collisions there is no such strong correlation.

\begin{table*}[htb]
\centering
\resizebox{\columnwidth*2}{!}{
			\begin{tabular}{l|l|l|rrrr|rrrr|rrrr|}
			 & & & \multicolumn{4}{c|}{$N_{\mathrm{jets}}(\%)$, All} & \multicolumn{4}{c|}{$N_{\mathrm{jets}}(\%)$, pencil-like} & \multicolumn{4}{c|}{$N_{\mathrm{jets}}(\%)$, sphere-like}\\
			Data Set	& energy & MPI							& 0 & 1 & 2 & 3+ & 0 & 1 & 2 & 3+ & 0 & 1 & 2 & 3+ \\ \midrule
			$e^{+}e^{-}$ & $91~$GeV & $-$				& $16.1$ & $11.8$ & $44.9$ & $27.2$ & $1.9$ & $5.9$ & $83.8$ & $8.4$ & $58.8$ & $11.0$ & $10.4$ & $19.8$ \\
			pp & $200~$GeV & ON					& $>99.9$ & $<0.01$ & $<0.01$ & $<0.01$ & $>99.9$ & $<0.01$ & $<0.01$ & $<0.01$ & $>99.9$ & $<0.01$ & $<0.01$ & $<0.01$ \\
			pp & $200~$GeV & OFF 					& $>99.9$ & $<0.01$ & $<0.01$ & $<0.01$ & $>99.9$ & $0.02$ & $<0.01$ & $<0.01$ & $>99.9$ & $<0.01$ & $<0.01$ & $<0.01$ \\
			pp & $13~$TeV & ON 					& $85.5$ & $9.2$ & $3.2$ & $2.1$ & $93.1$ & $5.5$ & $1.3$ & $0.2$ & $83.9$ & $8.7$ & $4.4$ & $3.0$ \\
			pp & $13~$TeV & OFF 				& $93.9$ & $5.2$ & $0.8$ & $0.1$ & $83.6$ & $13.9$ & $2.2$ & $0.3$ & $99.2$ & $0.6$ & $0.1$ & $0.1$ \\ \bottomrule
			\end{tabular}
			}
			\caption{Percentage of events containing $N_{\mathrm{jets}}=0,1,2,3+$ jets with $E>10~$GeV, for ensembles of all events, the $10\%$ most pencil-like events, and $10\%$ most sphere-like events as calculated using the transverse sphericity $S_{\mathrm{T}}$.}
			\label{tab:EventShape_jets_fractions_10}
			
\end{table*}

\begin{figure*}[htb]
\centering
  \includegraphics[width=0.44\linewidth]{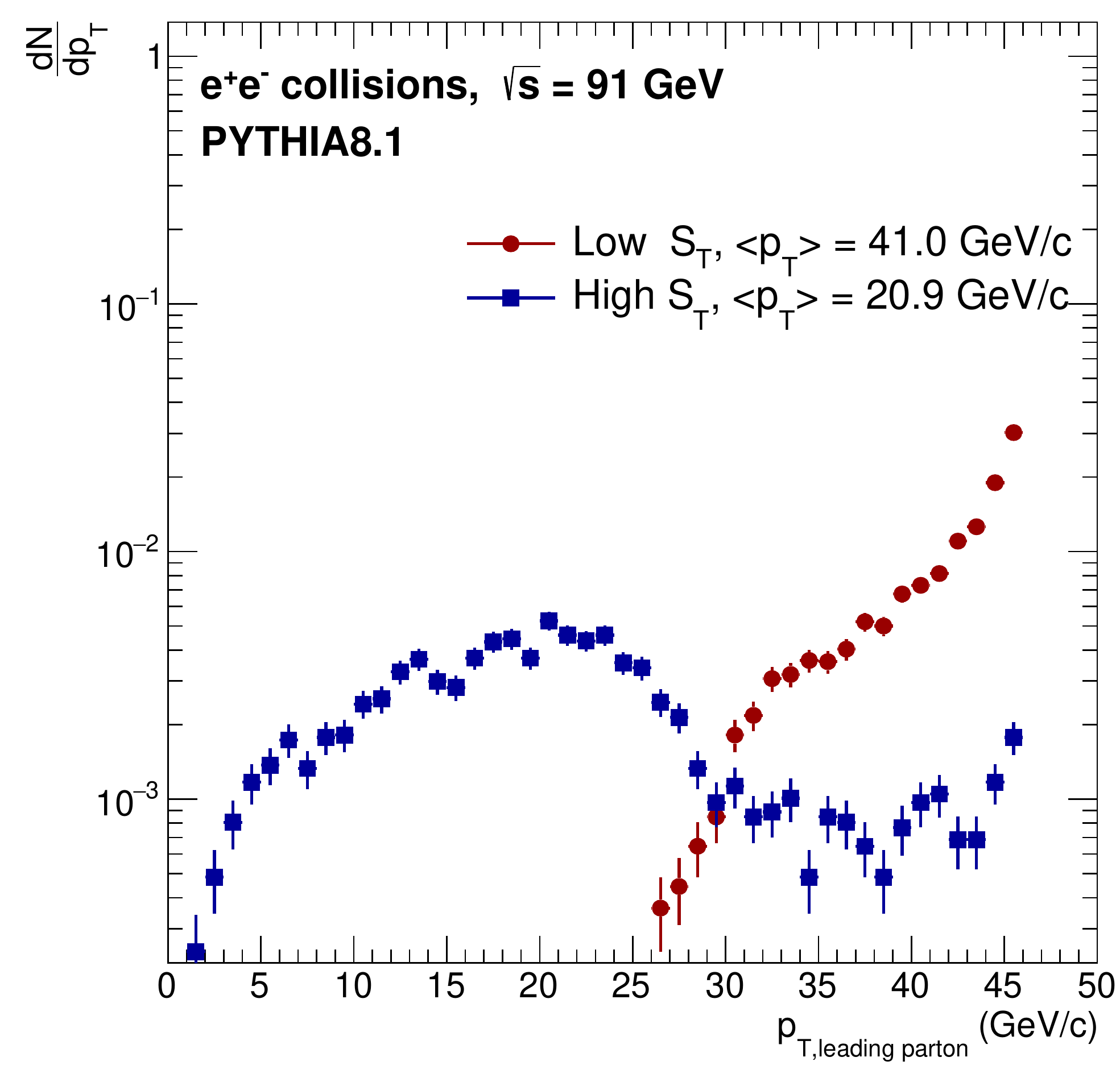}\\
  \includegraphics[width=0.44\linewidth]{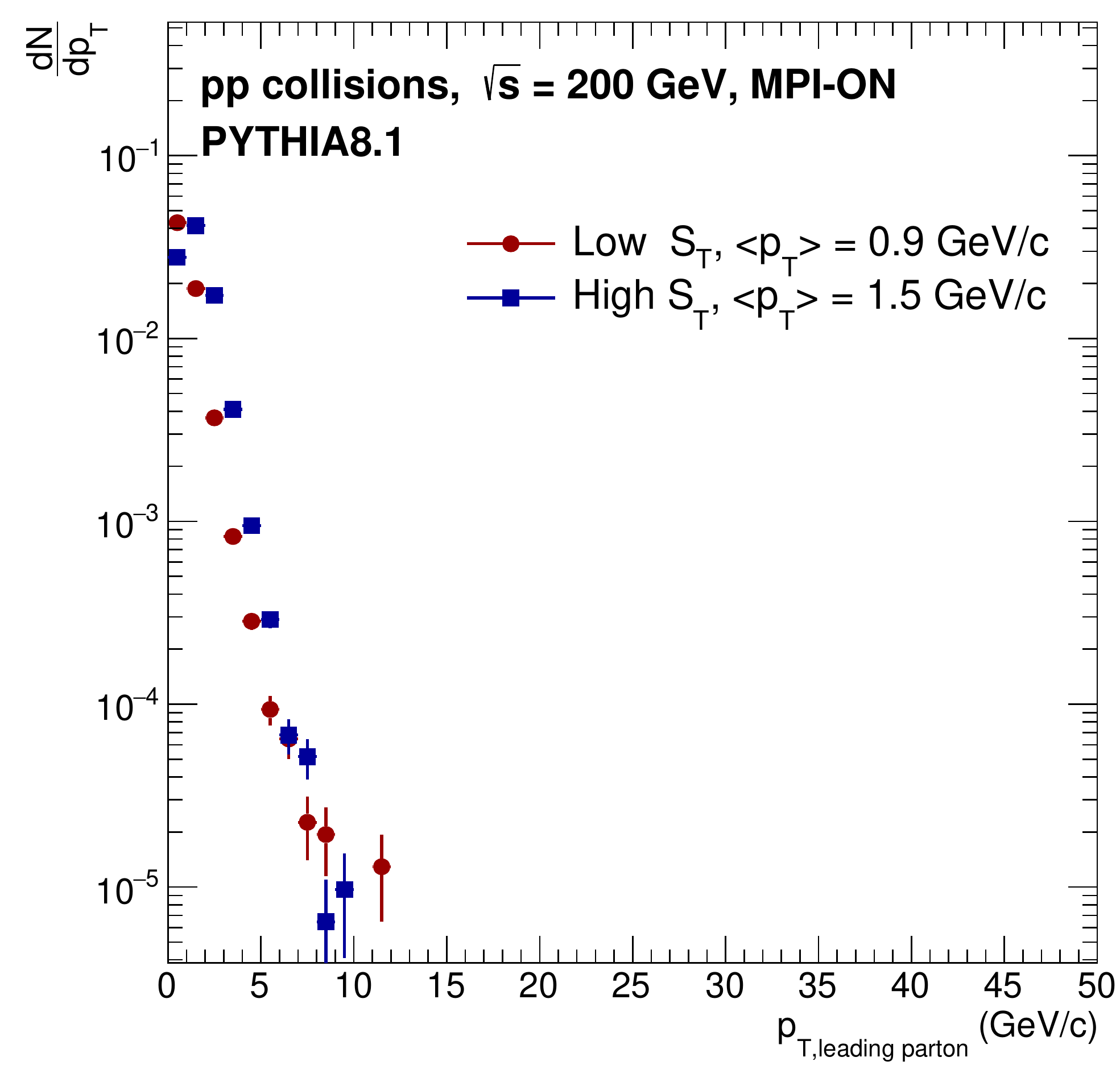}
  \includegraphics[width=0.44\linewidth]{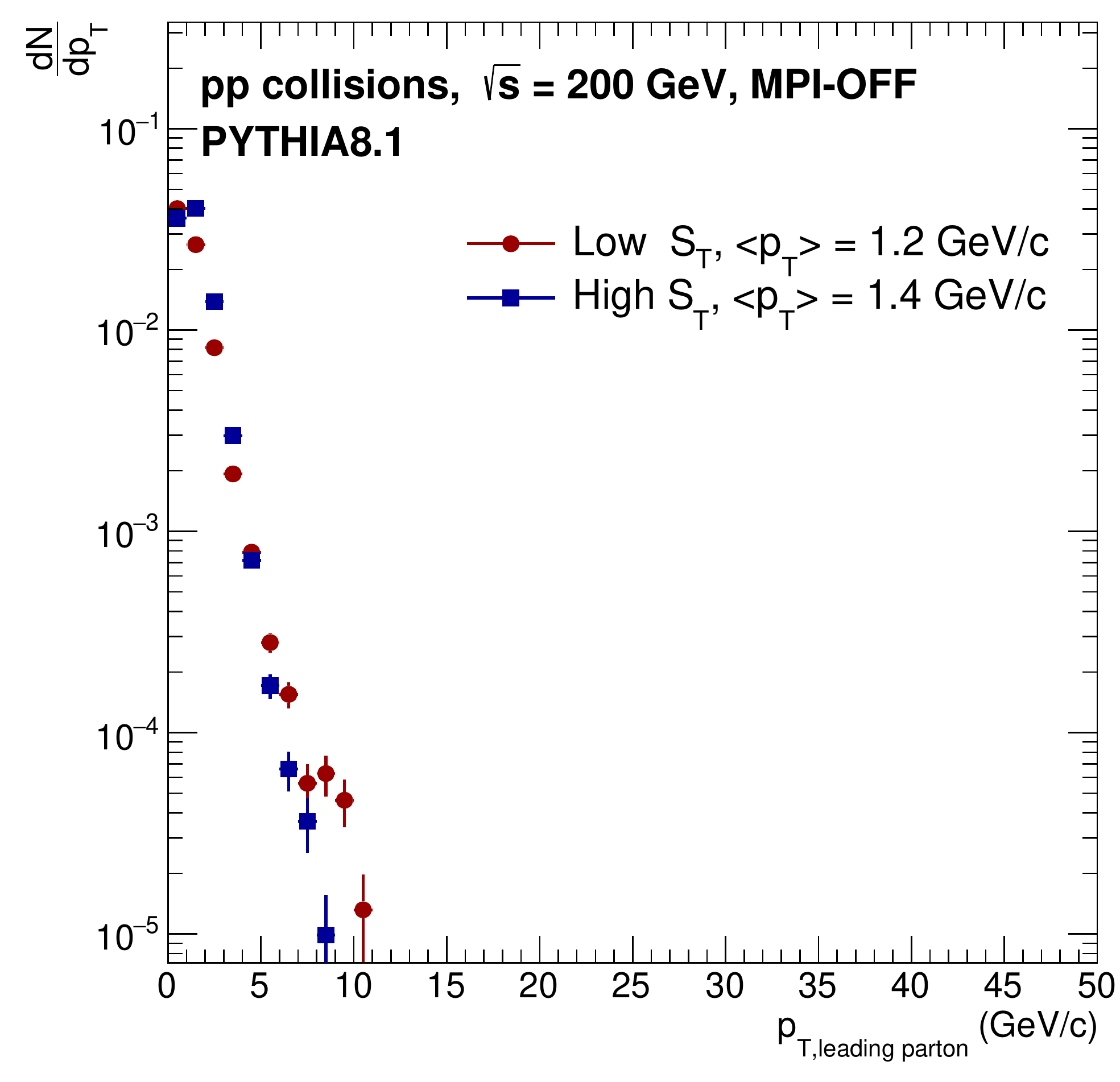}\\
  \includegraphics[width=0.44\linewidth]{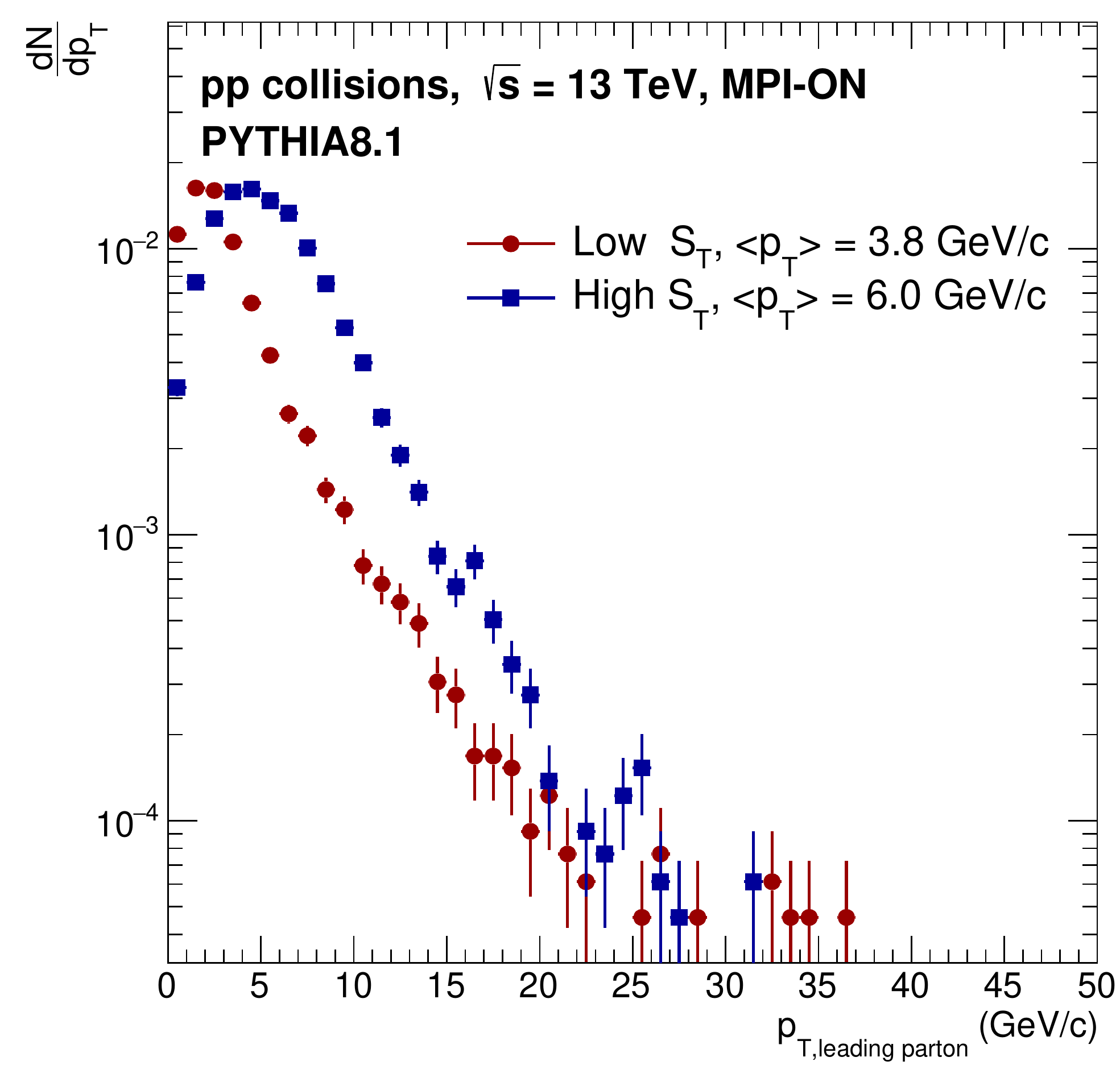}
  \includegraphics[width=0.44\linewidth]{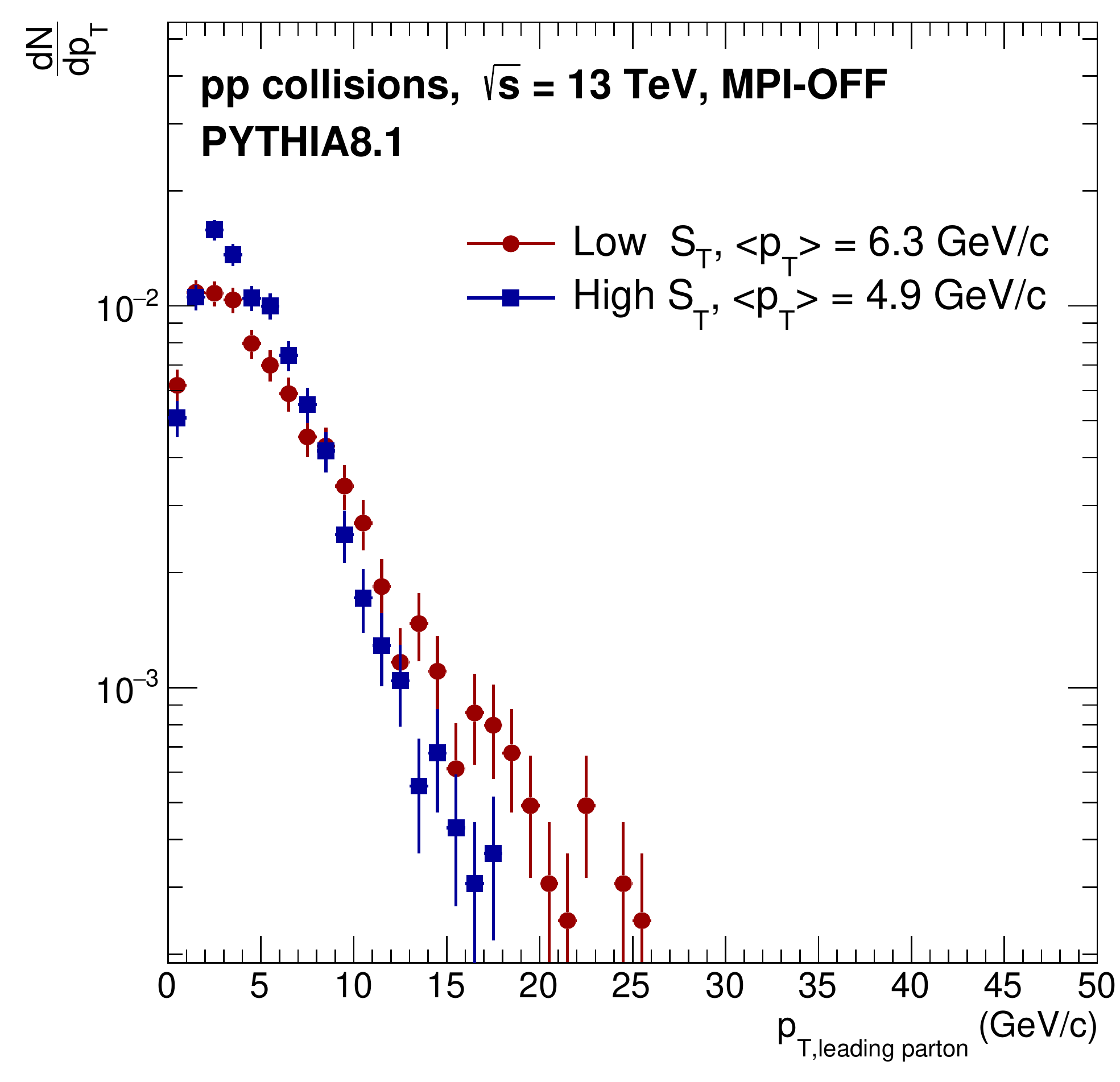}
  \caption{The distributions for the $\pT$ of the leading parton for pencil-like (low $S_{\mathrm{T}}$) and sphere-like (high $S_{\mathrm{T}}$) events, as calculated for each respective dataset. The mean of each distribution is given in the figure.}
  \label{fig:EventObservablesLowSHighS}
\end{figure*}

Now, since the correlation between the shape of the event and the number of reconstructed jets in \pp collisions is rather weak, it would be interesting to establish if there is another event characteristic that correlates well with the event shape. The distributions of the $\pT$ of the leading parton within the event for pencil-like and sphere-like events for $e^{+}e^{-}$, pp (MPI-OFF), and pp (MPI-ON) collisions are shown in Fig. \ref{fig:EventObservablesLowSHighS}. Similar to the previous results, the $10\%$ most pencil-like events (low $S_{\mathrm{T}}$) and $10\%$ most sphere-like events (high $S_{\mathrm{T}}$) are used to obtain the results. It indicates that selecting pencil- rather than sphere-like events leads to an increase of the mean $\pT$ of the leading parton of $\sim100\%$ for $e^{+}e^{-}$ collisions. This is consistent with the previous results presented in Fig. \ref{fig:EventShape_all}, as multi-jet topologies correlate with higher values of $S_{\mathrm{T}}$. Interestingly, for pp collisions, the mean $\pT$ of the leading parton is larger for sphere-like events compared to pencil-like events for the case with MPI-ON, while it is not the case with MPI-OFF. This observation is consistent with the conclusions taken from Table \ref{tab:EventShape_jets_fractions_10}.

The distributions for the number of multi-parton interactions for \pp collisions (MPI-ON) are shown in Fig. \ref{fig:EventObservablesLowSHighS_ppMPIon}, for pencil- and sphere-like events (left), as well as for the events with the $10\%$ lowest and highest final state particle multiplicities (right). It shows that events with higher sphericity and higher multiplicity both increase the mean number of multi-parton interactions, which can be understood from the idea that sphericity correlates with multiplicity. Moreover, $N_{\mathrm{MPI}}$ increases more when selecting the events with highest multiplicity instead of the most sphere-like configuration, and also more strongly excludes lower $N_{\mathrm{MPI}}$. Thus, the multiplicity of the event is found to be a better observable to select events with a larger number of multi-parton interactions in \pp collisions.

\section{Conclusions}

The analysis presented in this paper investigates the relationship between the shape of the event, the number of reconstructed jets, and other event characteristics such as the particle multiplicity, $\pT$ of the leading parton, and the number of multi-parton interactions. In Fig. \ref{fig:EventShape_all} it is shown that the $S_{\mathrm{T}}$ distributions are qualitatively different for $e^{+}e^{-}$ and \pp collisions, as most $e^{+}e^{-}$ collisions produce a pencil-like shape, while the particles produced in a \pp collisions are on average in a more sphere-like configuration. Also, it turns out that a large percentage of $e^{+}e^{-}$ collisions contain a reconstructed jet, while this is not the case for \pp collisions. The correlation between the transverse sphericity $S_{\mathrm{T}}$ and the number of reconstructed jets in $e^{+}e^{-}$ collisions enables the use of this observable to discriminate between di-jet and multi-jet topologies. Surprisingly, it turns out that in \pp collisions (MPI-ON) $S_{\mathrm{T}}$ is not strongly correlated to $N_{\mathrm{jets}}$, as any selection in $S_{\mathrm{T}}$ results in a sample dominated by $N_{\mathrm{jets}}=0$. Thus, these results suggest that \pp collisions with a pencil-like shape are by no means more jet-like compared to events without a selection on their shape. 

\begin{figure*}[htb]
\centering
  \includegraphics[width=0.44\linewidth]{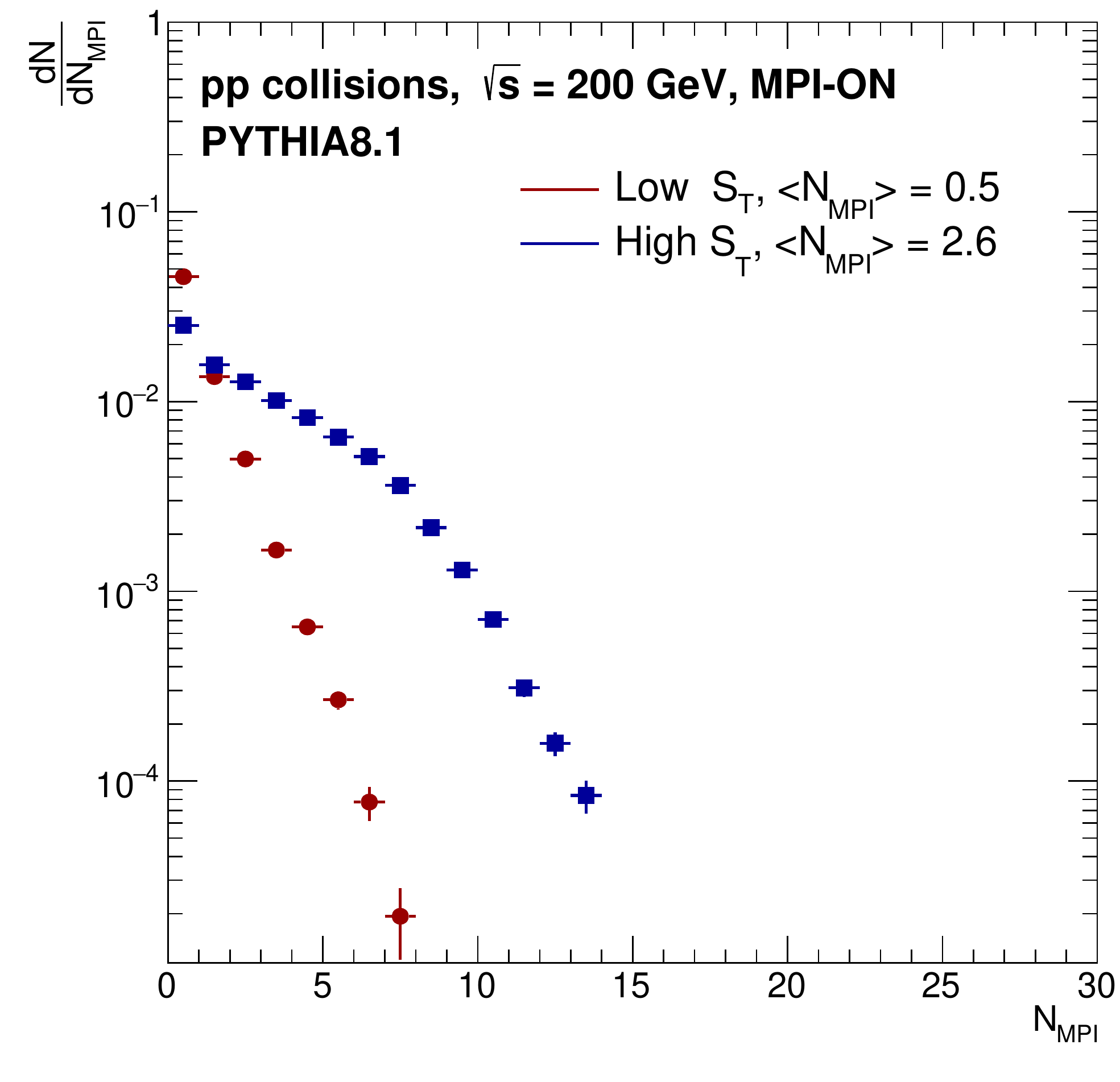}
  \includegraphics[width=0.44\linewidth]{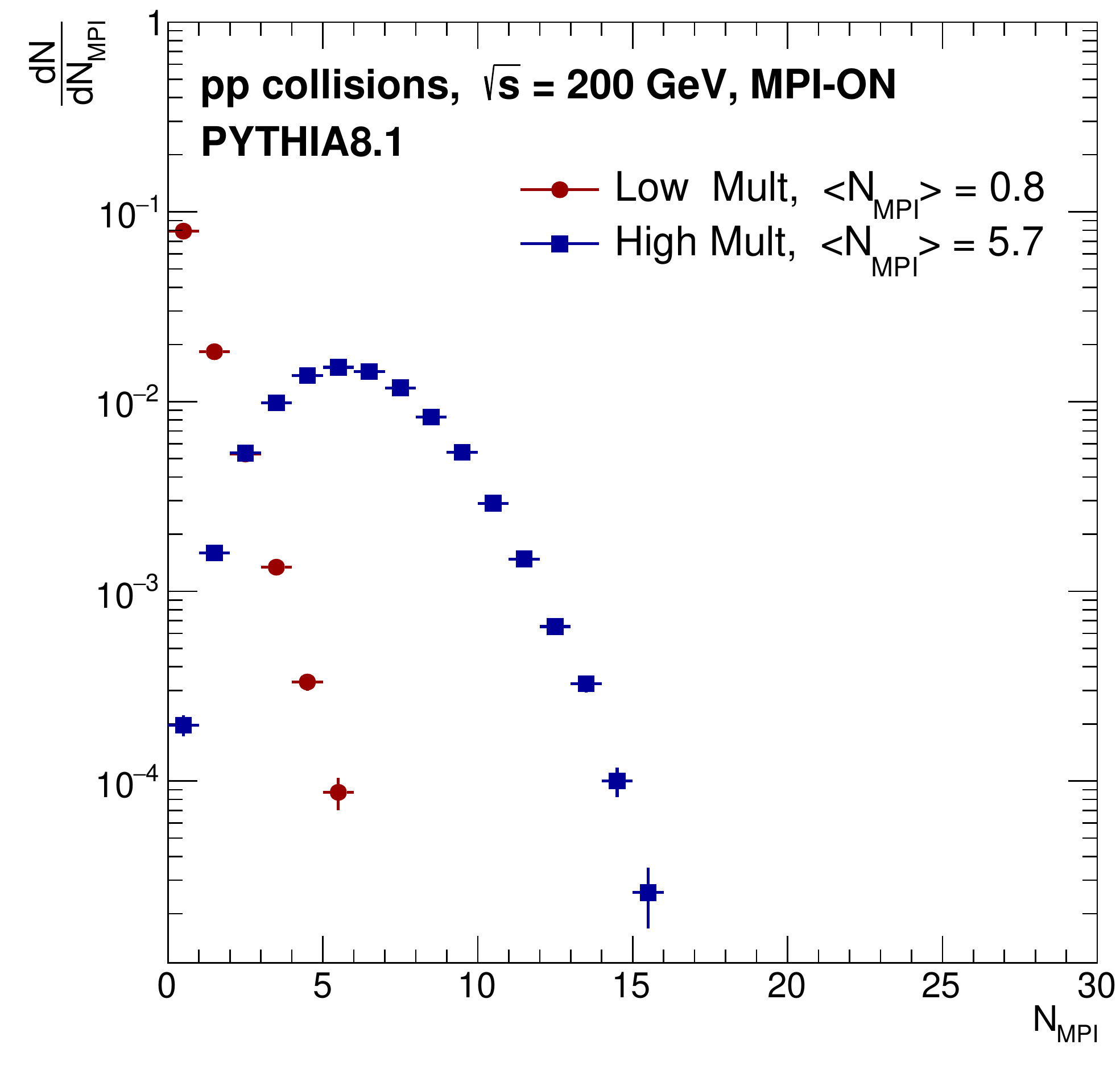}\\
  \includegraphics[width=0.44\linewidth]{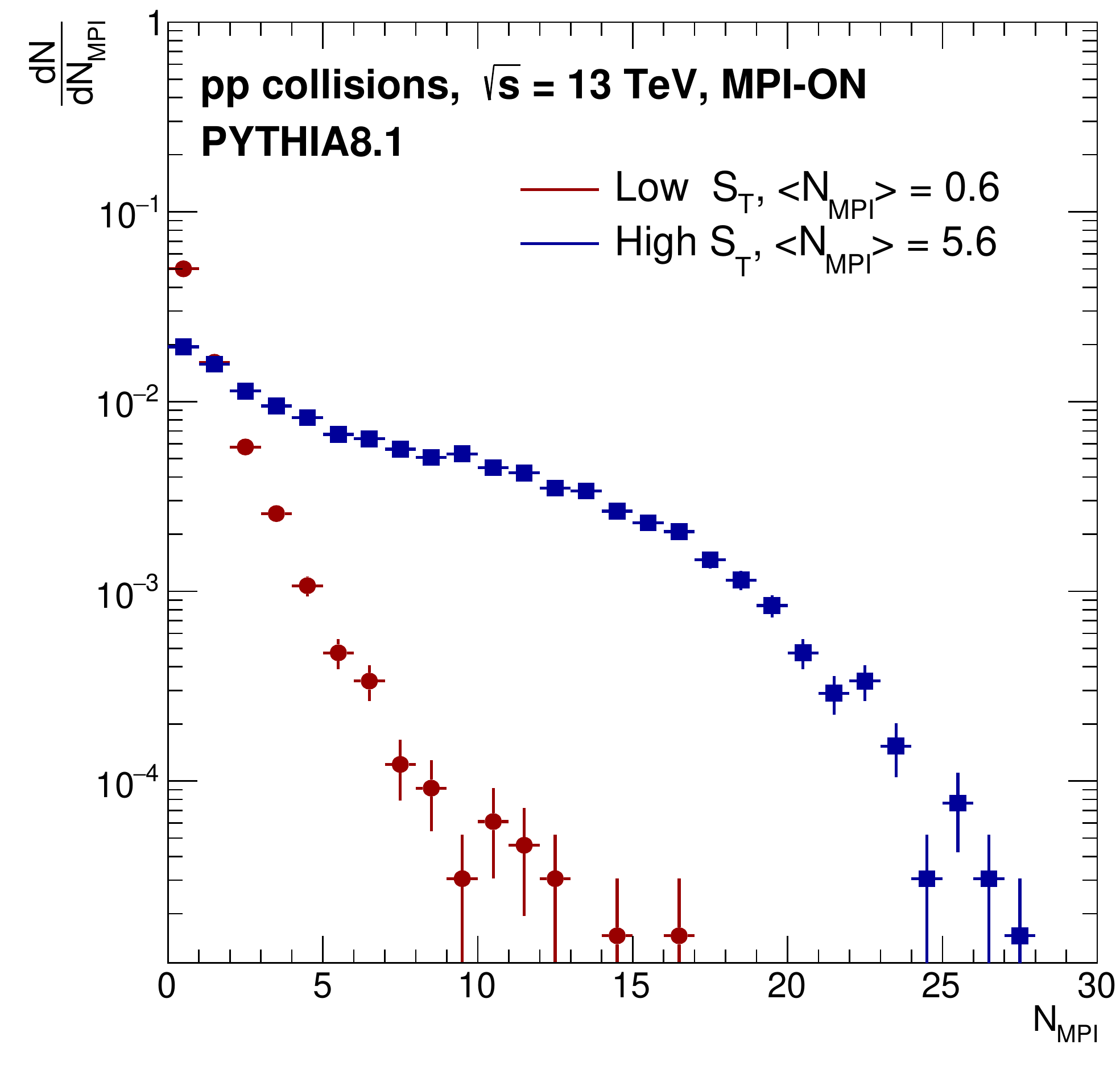}
  \includegraphics[width=0.44\linewidth]{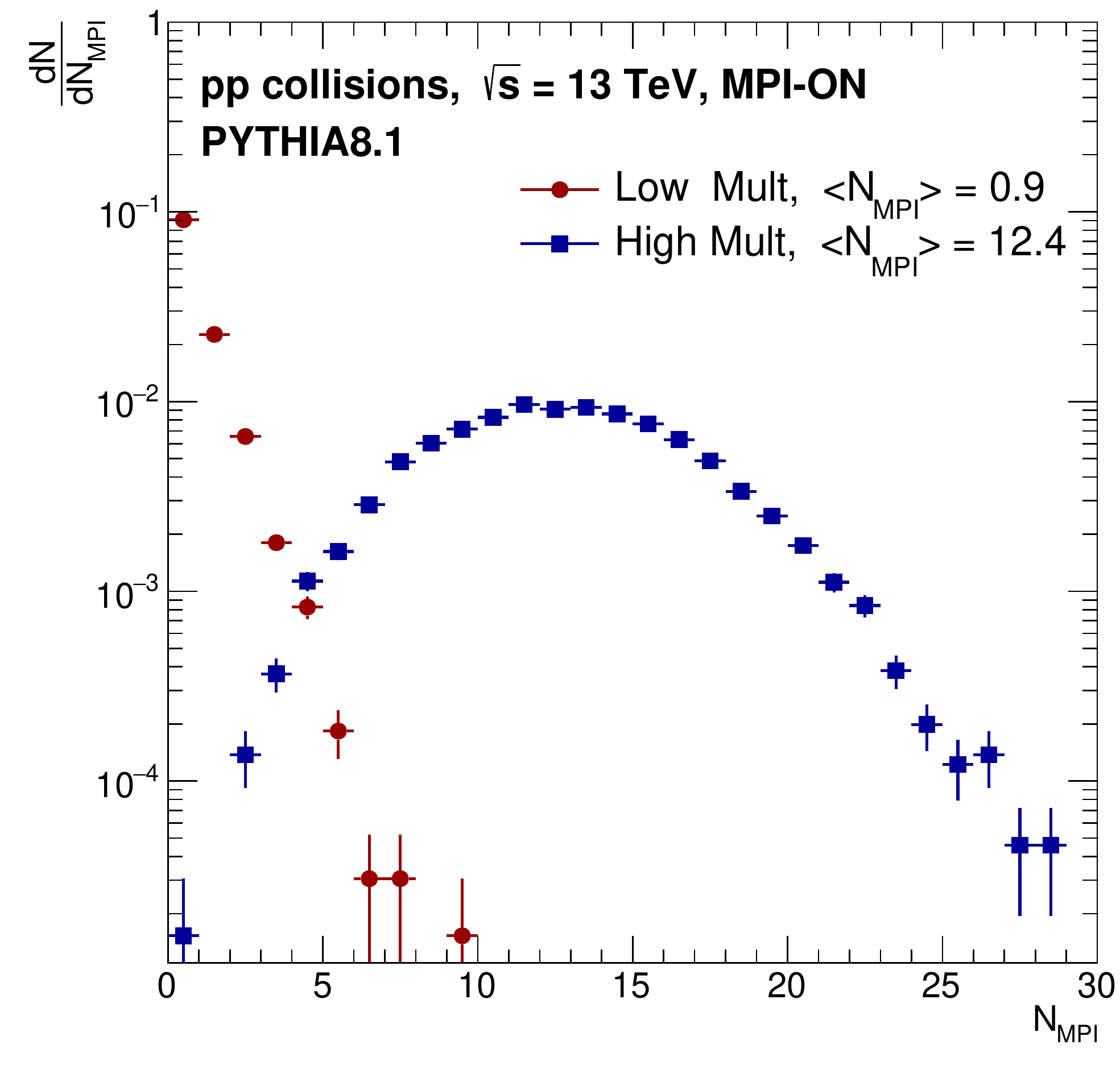}
  \caption{The distributions of the number of multi-parton interactions for pencil- and sphere-like events (left), as well as events with low and high multiplicity (right), as calculated for simulated \pp (MPI-ON) collisions at $\sqrt{s}=200~$GeV (top) and $\sqrt{s}=13~$TeV (bottom), respectively. The mean of each distribution is given in the figure.}
  \label{fig:EventObservablesLowSHighS_ppMPIon}
\end{figure*}

Furthermore, the results shown in Fig. \ref{fig:EventObservablesLowSHighS_ppMPIon} indicate a strong correlation between $S_{\mathrm{T}}$ and the number of multi-parton interactions. The average number of these interactions increases relatively more for pp collisions at $\sqrt{s}=13~$TeV compared to pp collisions at $\sqrt{s}=200~$GeV. However, as Fig. \ref{fig:EventObservablesLowSHighS_ppMPIon} (right) shows, the mean number of multi-parton interactions increase even more between events with low and high multiplicities. These observations are consistent with the idea that $S_{\mathrm{T}}$ correlates with the particle multiplicity, i.e. a large number of multi-parton interactions lead to a more sphere-like configuration with increased multiplicity. As it is experimentally impossible to directly measure the $N_{\mathrm{MPI}}$, it would be very interesting to use $S_{\mathrm{T}}$, as well as the event multiplicity, to explore \pp collisions with small and large amounts  of multi-parton interactions and compare them to model predictions.

\section{ACKNOWLEDGEMENTS}
This work was supported by the Netherlands Organisation for Scientific Research (NWO). This work was supported in part by the Office of Nuclear Physics of the U.S. Department of Energy. Work supported by the US DOE under award number DE-SC004168.

%
%
\ifboolexpr{bool{jacowbiblatex}}%
	{\printbibliography}%
	{%
	

}

%
%


\end{document}